\documentclass[submission, Proceedings]{SciPost}

\hypersetup{
    colorlinks,
    linkcolor={red!50!black},
    citecolor={blue!50!black},
    urlcolor={blue!80!black}
}

\usepackage[bitstream-charter]{mathdesign}
\DeclareSymbolFont{usualmathcal}{OMS}{cmsy}{m}{n}
\DeclareSymbolFontAlphabet{\mathcal}{usualmathcal}

\begin{document}
%\linenumbers

\begin{center}{\Large \textbf{
 Longitudinal and Transverse Spin Transfer of $\Lambda$ and $\overline{\Lambda}$ Hyperons in Polarized $p$+$p$ Collisions at $\sqrt{s} = 200$ GeV at RHIC-STAR\\
}}\end{center}

% TODO: write the author list here. Use initials + surname format.
% Separate subsequent authors by a comma, omit comma at the end of the list.
% Mark the corresponding author with a superscript *.
\begin{center}
Yike Xu\textsuperscript{1 *}

for STAR Collaboration

\end{center}

\begin{center}
{\bf 1} Institute of Frontier and Interdisciplinary Science \& Key Laboratory of Particle Physics and Particle Irradiation (MOE), 
Shandong University, Qingdao, Shandong, 266237, China
\\
%\institute{ Institute of Frontier and Interdisciplinary Science \&  Key Laboratory of Particle Physics and Particle Irradiation of Minstry of Education,  
%  Shandong University, Qingdao, Shandong, 266237, China}
* yxu@rcf.rhic.bnl.gov
\end{center}

\begin{center}
\today
\end{center}

% For convenience during refereeing (optional),
% you can turn on line numbers by uncommenting the next line:
%\linenumbers

\definecolor{palegray}{gray}{0.95}
\begin{center}
\colorbox{palegray}{
  \begin{tabular}{rr}
  \begin{minipage}{0.1\textwidth}
    \includegraphics[width=22mm]{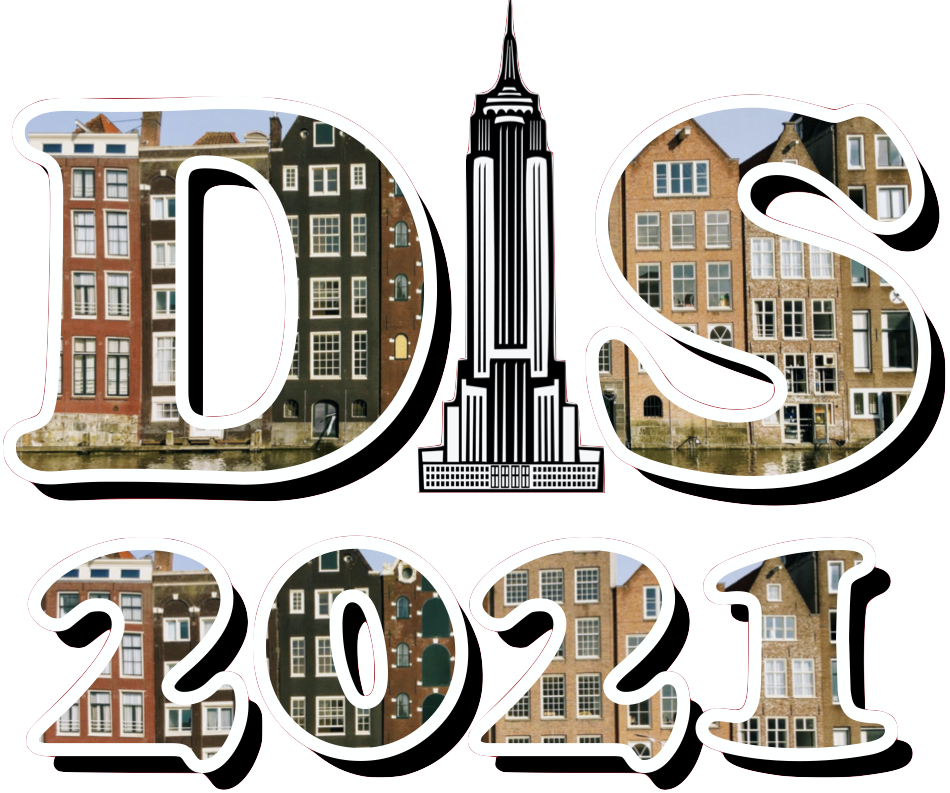}
  \end{minipage}
  &
  \begin{minipage}{0.75\textwidth}
    \begin{center}
    {\it Proceedings for the XXVIII International Workshop\\ on Deep-Inelastic Scattering and
Related Subjects,}\\
    {\it Stony Brook University, New York, USA, 12-16 April 2021} \\
    \doi{10.21468/SciPostPhysProc.?}\\
    \end{center}
  \end{minipage}
\end{tabular}
}
\end{center}

\section*{Abstract}
{\bf
% TODO: write your abstract here.

Measurements of the longitudinal spin transfer, $D_{LL}$, and the transverse spin transfer, $D_{TT}$, of the $\Lambda(\bar{\Lambda}$) hyperon in polarized $p$+$p$ collisions are expected to be sensitive to the helicity distribution and transversity distribution of the $s(\bar{s})$ quark in proton and the corresponding polarized fragmentation functions.
This contribution presents the new preliminary results of the $\Lambda(\bar{\Lambda}$) $D_{LL}$ and $D_{TT}$ using data collected at RHIC-STAR experiment in 2015, with twice larger statistics than previously published results.
}


%\vspace{10pt}
%\noindent\rule{\textwidth}{1pt}
%\tableofcontents\thispagestyle{fancy}
%\noindent\rule{\textwidth}{1pt}
%\vspace{10pt}

%Introduction
\section{Introduction}
\label{sec:intro}
Since the surprising results on the spin structure of the proton by the EMC experiment in the late 1980s~\cite{EMC}, 
much progress has been made in understanding the origin of the proton spin. 
However, the sea quark contribution to the proton spin, for example, the polarized distributions of the strange quark(anti-quark), $s(\bar{s})$, 
is still not well constrained by experimental data. 
The $\Lambda$ hyperon contains a strange constitute quark, which is expected to carry most of $\Lambda$ spin. The  $\Lambda$ polarization, $P_{\Lambda(\bar{\Lambda})}$ , can be determined from the angular distribution of the weakly decayed daughters. 
The spin transfer of the $\Lambda$ hyperon in proton-proton collisions provides a way to study the $s(\bar{s})$ quark contribution to the proton spin~\cite{DLL3se, Qinghua2004,Qinghua2006dtt}. 
Recently, strange quark polarization has been extracted from a model calculation incorporating the STAR measurements of the hyperon spin transfer~\cite{xiaonan}.

The longitudinal and transverse spin transfer asymmetries, $D_{LL}$ and $D_{TT}$, of $\Lambda$ hyperons in $p$+$p$ collisions are defined in Eq \ref{defination}. They are naturally connected to the polarized parton distribution functions in the proton and polarized fragmentation functions.
$D_{LL}$ provides access to the helicity distribution of strange quark, while $D_{TT}$ is coupled to the transversity distribution.
\begin{gather}
D_{LL}^{\Lambda} \equiv \frac{d\sigma^{(p^{+}p \to \Lambda^{+}X)} - d\sigma^{(p^{+}p \to \Lambda^{-}X)}}{d\sigma^{(p^{+}p \to \Lambda^{+}X)} + d\sigma^{(p^{+}p \to \Lambda^{-}X)}} = \frac{d\Delta\sigma^{\Lambda}}{d\sigma^{\Lambda}}; \notag \\
D_{TT}^{\Lambda} \equiv \frac{d\sigma^{(p^{\uparrow}p \to \Lambda^{\uparrow}X)} - d\sigma^{(p^{\uparrow}p \to \Lambda^{\downarrow}X)}}{d\sigma^{(p^{\uparrow}p \to \Lambda^{\uparrow}X)} + d\sigma^{(p^{\uparrow}p \to \Lambda^{\downarrow}X)}} = \frac{d\delta\sigma^{\Lambda}}{d\sigma^{\Lambda}},  
\label{defination}
\end{gather}
where $p^+$/$p^-$ and $\Lambda^+$/$\Lambda^-$ denote the helicity of the colliding proton and the $\Lambda$ hyperon, and $\uparrow$/$\downarrow$ denotes the positive or negative transverse polarization of them. $d\delta\sigma^{\Lambda}$ ($d\Delta \sigma^{\Lambda}$) is the transversely (longitudinally) polarized cross section and $d\sigma^{\Lambda}$ is the unpolarized cross section. The polarized cross section can be factorized into the convolution of parton distribution functions, polarized partonic cross-section and the polarized fragmentation function.

This contribution presents the new preliminary results of $D_{LL}$ and $D_{TT}$ for $\Lambda(\bar{\Lambda})$ in proton-proton collisions at $\sqrt{s} = 200$ GeV, 
with hyperon pseudo-rapidity $|\eta| < 1.2$ and transverse momenta up to $8.0$ $\rm{GeV}/$$c$. 
This dataset is about twice as large as the previously published $D_{LL}$ and $D_{TT}$ results~\cite{AdamJimprovedDLLMeasurement, AdamJimprovedDTTMeasurement}.

%Experiments
\section{Experiments and Hyperon Reconstruction}
The Relativistic Heavy Ion Collider (RHIC) is the world’s first and only polarized hadron-hadron collider. RHIC runs proton-proton collisions at $\sqrt{s} = 200$ GeV and $\sqrt{s} = 510$ GeV with proton beams longitudinally or transversely polarized. In 2015, RHIC delivered 200 GeV $p$+$p$ collisions with an average beam polarization of $56\%$ and an integrated luminosity of $52$ $\rm{pb^{-1}}$ .
The STAR (Solenoidal Tracker At RHIC) experiment~\cite{STAR} is located at the 6 o'clock position of the RHIC ring. For $D_{TT} $ and $D_{LL}$ analyses, several sub-detectors were used, including the Time Projection Chamber (TPC), ElectroMagnetic Calorimeter (EMC), and Time Of Flight (TOF) detector.
Hard scattering events were selected with a Jet Patch trigger which was based on energy depositions in the EMC. Additionally, BBC (Beam-Beam Counter) and VPD (Vertex Position Detector) were used to monitor the relative luminosity.

\begin{figure}
    \centering
    \includegraphics[width=0.49\columnwidth]{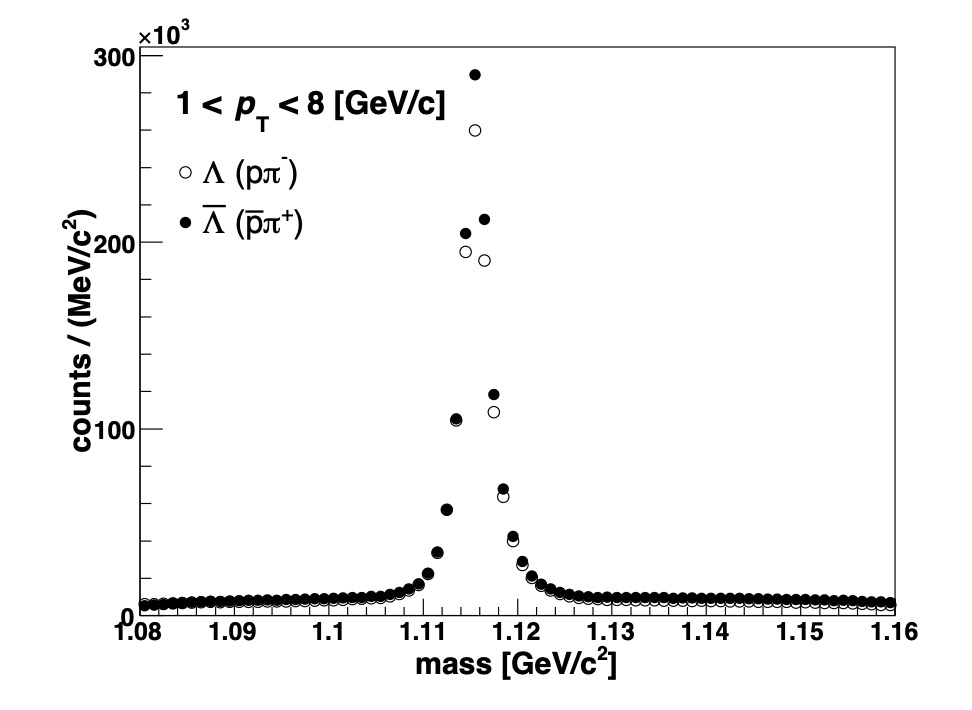}
\captionsetup{width=0.99\columnwidth}
    \caption{The invariant mass distribution for $\Lambda$ and $\bar{\Lambda}$ candidates with 1 < $p_T$ < 8 GeV/$c$ after applying selection cuts as in~\cite{AdamJimprovedDTTMeasurement}. }
    \label{fig:lambda}
\end{figure}

The $\Lambda(\bar{\Lambda})$ candidates are identified from the topology of their weak decay channels, $\Lambda\to p\pi^-$ ($\bar{\Lambda} \to \bar{p}\pi^+$). Pairs of proton and pion tracks measured in the TPC are used to reconstruct the $\Lambda(\bar{\Lambda})$ candidates. TOF information is also used to improve particle identification. Then, a series of topological cuts are tuned to further reduce the background. The side-band method is used as in~\cite{AdamJimprovedDLLMeasurement,AdamJimprovedDTTMeasurement} to estimate the residual background fraction, which is less than 10\%. 
The spin transfer of the $\Lambda$ hyperon is obtained by subtracting the contribution from residual background, for example, $D_{LL} = (D_{LL}^{raw}-rD_{LL}^{bkg})/(1-r)$, where $D_{LL}^{raw}$ and $D_{LL}^{bkg}$ are the spin transfers for signal and side-band regions, and $r$ is the background fraction. A similar definition is used for $D_{TT}$. Figure~\ref{fig:lambda} shows an example of $\Lambda$ candidates invariant mass distribution.

The spin transfer analyses are performed for the $\Lambda(\bar{\Lambda})$ candidates associated with a jet.  The reconstructed jet axis is used as a proxy of the momentum direction of the fragmenting quark, which is needed in determining the polarization direction for $D_{TT}$ measurements~\cite{AdamJimprovedDTTMeasurement}. The anti-$k_T$ algorithm is used to reconstruct jets, similar as in Ref.~\cite{jetrec1,STAR:2019yqm}, with resolution parameter of $R$ = 0.6. Then, we require jet transverse momentum $p_T>5$ $\rm{GeV}$/$c$ and $-0.7<\eta_{detector}<0.9$. The detector pseudorapidity, $\eta_{detector}$, is defined by extrapolating the jet thrust axis into the BEMC detector, and calculating the pseudorapidity of that intersection point relative to the center of the STAR detector. The correlation between $\Lambda(\bar{\Lambda})$ candidates and the reconstructed jets is made by constraining the distance $\Delta R=\sqrt{(\Delta\eta)^2+(\Delta\phi)^2 }$ between momentum directions of $\Lambda(\bar{\Lambda})$ candidates and the reconstructed jets in $\eta-\phi$ space. The $\Lambda(\bar{\Lambda})$ hyperons in the near-side of jets $(\Delta R<0.6)$ are kept for the $D_{TT}$ and $D_{LL}$ analyses.

%Results
\section{$D_{LL} $ \& $D_{TT}$ Measurements and Results}

%subsection $D_{LL}$ measurements 
\subsection{$D_{LL}$ Measurements and Results}
$D_{LL}$ is measured from asymmetry of $\Lambda$ counts with opposite beam polarizations in small $\cos\theta^*$ bins, where the $\theta^*$ is the angle between the $\Lambda(\bar{\Lambda})$ polarization direction and the (anti-) proton momentum in the $\Lambda(\bar{\Lambda})$ rest frame~\cite{DLLMeasurement}:
\begin{equation}
 D_{LL} = \frac{1}{\alpha P_{beam} \left \langle \cos\theta^* \right \rangle } \frac{N^{+}-RN^{-}}{N^{+}+RN^{-}}
\end{equation}
where the $\left \langle \cos\theta^* \right \rangle$ is the average value of each $\cos\theta^*$ bin, $N^{+/-}$ is the $\Lambda$ count in a $\cos\theta^*$ bin when the helicity of the polarized beam is positive or negative, $R = \mathscr L^{+} / \mathscr L^{-}$ is the relative luminosity ratio, $\alpha=0.732\pm0.014$ is the decay parameter and $P_{beam}$ is the beam polarization.

\begin{figure}
    \centering
    \includegraphics[width=0.55\columnwidth]{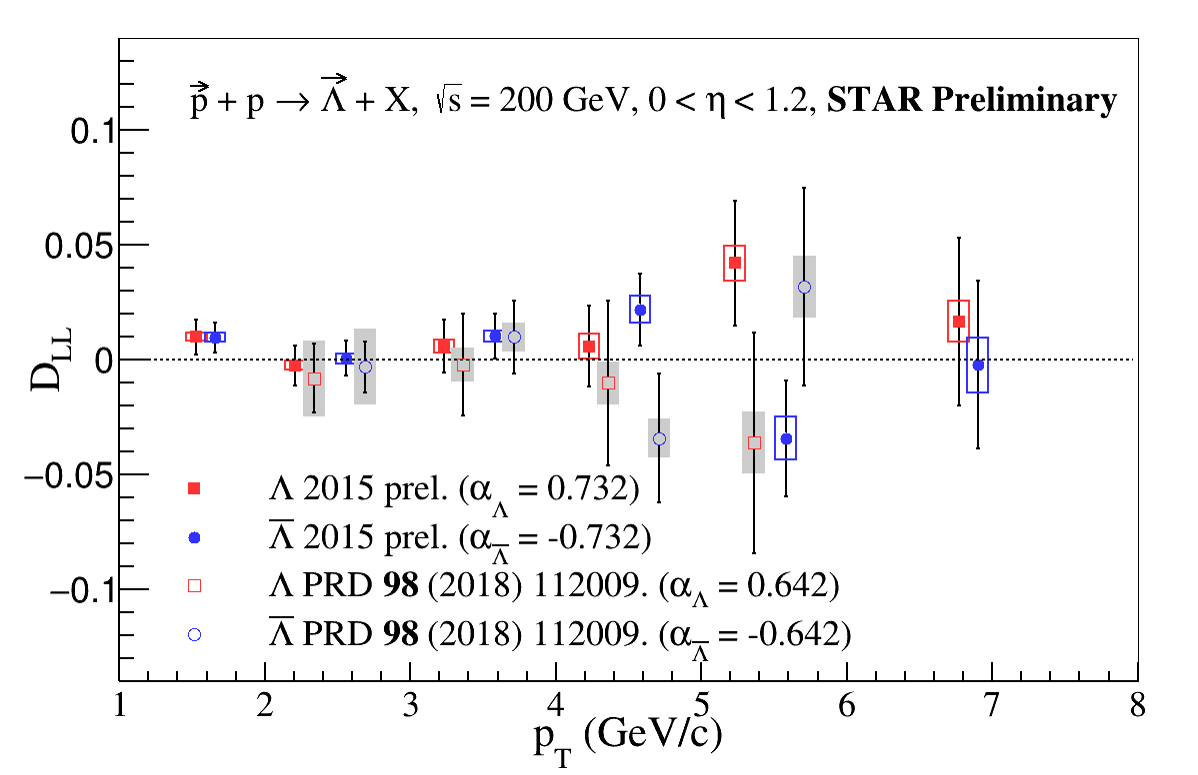}
    \captionsetup{width=0.99\columnwidth}
    \caption{Preliminary results of $D_{LL}$ for $\Lambda$ and $\bar{\Lambda}$ from STAR 2015 data versus hyperon $p_T$, together with previously published results. The results for the $\bar{\Lambda}$ and the previously published results have been shifted slightly to larger $p_T$ for clarity.}
    \label{fig:figure_LambdDLL_vs_pT}
\end{figure}

Figure \ref{fig:figure_LambdDLL_vs_pT} shows new $D_{LL}$ preliminary results based on the 2015 data at 200 GeV, together with the  published results from the 2009 data~\cite{AdamJimprovedDLLMeasurement}. The new results have about two times larger statistics than the previously published results, and are consistent with them. The new results cover transverse momenta up to $8.0$ $\rm{GeV}/$$c$, and are consistent with zero within uncertainties.  These results are also comparable with model calculations in~\cite{DLL3se}.  
%These new data can also be used to constrain the $\Lambda$ polarized fragmentation functions.
%The small spin transfer is probably expected as the recent global analysis indicates a smaller strange helicity distribution. 

%subsection $D_{TT}$ measurements
\subsection{$D_{TT}$ Measurements and Results}

%For the transverse spin transfer $D_{TT}$, the transverse polarization direction of the outgoing parton after partonic scattering is used as the hyperon polarization direction. The rotation along the normal direction of the partonic scattering plane spanned by the transverse polarization directions of the incoming and outgoing quarks is considered for the $\Lambda$ hyperons polarization direction

For the transverse spin transfer $D_{TT}$, the hyperon polarization direction is defined as the transverse polarization direction of the outgoing parton, which is obtained by rotating the polarization vector of the incoming parton along the normal direction of the partonic scattering plane. The scattering plane is spanned by the momentum directions of incoming and outgoing partons~\cite{AdamJimprovedDTTMeasurement}. Here the reconstructed jet axis is used as the momentum direction of the outgoing parton. 

$D_{TT}$ is measured from a cross-ratio asymmetry using $\Lambda$ counts with opposite beam polarizations within small $\cos \theta^*$ bins~\cite{AdamJimprovedDTTMeasurement}:
\begin{equation} 
D_{TT} = \frac{1}{\alpha P_{beam} \left \langle \cos\theta^* \right \rangle } \frac{\sqrt{N^{\uparrow}(\cos \theta^*)N^{\downarrow}(-\cos \theta^*)}-\sqrt{N^{\uparrow}(-\cos\theta^*)N^{\downarrow}(\cos\theta^*)}}{\sqrt{N^{\uparrow}(\cos\theta^*)N^{\downarrow}(-\cos\theta^*)}+\sqrt{N^{\uparrow}(-\cos\theta^*)N^{\downarrow}(\cos\theta^*)}}.
\end{equation}
With the cross-ratio method, the relative luminosity and the acceptance are both cancelled, which helps to reduce systematic uncertainties. 

\begin{figure}
    \centering
    \includegraphics[width=0.61\columnwidth]{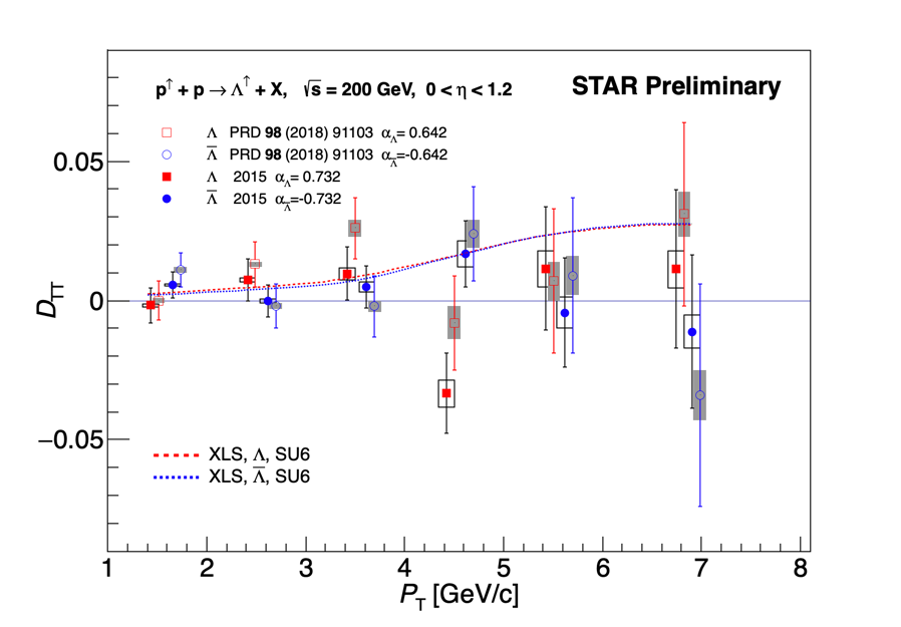}
    \captionsetup{width=0.99\columnwidth}
    \caption{Preliminary results of $D_{TT}$ for $\Lambda$ and $\bar{\Lambda}$ from STAR 2015 data versus hyperon $p_T$, together with previously published results.
The $\bar{\Lambda}$ results and the previously published results have been shifted slightly to larger $p_T$ values for clarity. }
    \label{fig:figure_DTT_vs_lambdapt}
\end{figure}

Figure~\ref{fig:figure_DTT_vs_lambdapt} shows the new preliminary $D_{TT}$ results from 2015 data, together with previously published results from 2012 dataset~\cite{AdamJimprovedDTTMeasurement} versus $\Lambda(\bar{\Lambda})$ $p_T$ in positive pseudo-rapidity region relative to the polarized beam. The statistical uncertainties of the new results exhibit a $\sqrt{2}$ improvement compared to previously published results as expected. The new results are consistent with zero within uncertainties, and are also consistent with model predictions~\cite{Qinghua2006dtt}.

%Summary and Outlook
\section{Summary and Outlook}
Longitudinal ($D_{LL}$) and transverse ($D_{TT}$) spin transfer measurements in $p$+$p$ collisions can provide access to the strange quark helicity and transversity distributions in the proton and the polarized fragmentation functions.
New preliminary results of $D_{LL}$ and $D_{TT}$ in $p$+$p$ collisions at 200 GeV from STAR 2015 dataset were reported, with twice the statistics of the previous results. The new results are consistent with previous measurements, and are consistent with zero within uncertainties, which indicate that the strange quark polarized distribution and/or the polarized fragmentation function of $\Lambda(\bar{\Lambda})$ is small. 

STAR has expanded its acceptance by installing a series of detector upgrades, in particular in the forward rapidity region. More proton-proton collision data will be collected at STAR in 2022 at 510 GeV and in 2024 at 200 GeV. That will significantly increase the statistics of $\Lambda$ hyperon samples. Those forward detector upgrades allow for a rich $\Lambda$ physics program in the forward rapidity region, including not only spin transfer measurements but also the transverse hyperon polarization in unpolarized proton collisions.

%%%%%%%%%%%%%%%%%%%%%%%%%%%%

\section*{Acknowledgements}
%Acknowledgements should follow immediately after the conclusion.

We thank the RHIC Operations Group and RCF at BNL.  The author is supported partially by the National Natural Science Foundation of China under No. 12075140.

\bibliography{SciPost_Example_BiBTeX_File.bib}

\begin{thebibliography}{99}

\bibitem{EMC}J. Ashman et al. (European Muon collaboration), {\it A measurement of the spin asymmetry and determination of the structure function g1 in deep inelastic muon-proton scattering}, Phys. Lett. {\bf B206}, 364 (1988),
\doi{10.1016/0370-2693(88)91523-7}.

\bibitem{DLL3se}D. de Florian, M. Stratmann, and W. Vogelsang, {\it Polarized $\Lambda$-Baryon Production in $pp$ Collisions}, Phys. Rev. Lett. {\bf 81}, 530 (1998),  
\doi{10.1103/PhysRevLett.81.530}.

\bibitem{Qinghua2004}Qing-hua Xu and Zuo-tang Liang, {\it Probing gluon helicity distribution and quark transversity through hyperon polarization in singly polarized $pp$ collisions}, Phys. Rev. D {\bf 70}, 034015 (2004), 
\doi{10.1103/PhysRevD.70.034015}.

\bibitem{Qinghua2006dtt}Qing-hua Xu, Zuo-tang Liang, and Ernst Sichtermann, {\it Anti-Lambda polarization in high energy $pp$ collisions with polarized beams}, Phys. Rev. D {\bf 73}, 077503 (2006), 
\doi{10.1103/PhysRevD.73.077503}.

\bibitem{xiaonan}Xiaonan Liu and Bo-Qiang Ma, {\it Nucleon strangeness polarization from $\Lambda /\bar{\Lambda }$ hyperon production in polarized proton–proton collision at RHIC}. Eur. Phys. J. C {\bf 79}, 409 (2019),
\doi{10.1140/epjc/s10052-019-6916-z}.

\bibitem{AdamJimprovedDLLMeasurement}Adam, Jaroslav et al. (STAR Collaboration), {\it Improved measurement of the longitudinal spin transfer to ${\Lambda}$ and $\bar{\Lambda}$ hyperons in polarized proton-proton collisions at $\sqrt{s} = 200$ ${GeV}$}, Phys. Rev. D {\bf 98}, 112009 (2018), \doi{10.1103/PhysRevD.98.112009}.

\bibitem{AdamJimprovedDTTMeasurement}Adam, Jaroslav et al. (STAR Collaboration), {\it Transverse spin transfer to ${\Lambda}$ and $\bar{\Lambda}$ hyperons in polarized proton-proton collisions at $\sqrt{s} = 200$ ${GeV}$}, Phys. Rev. D {\bf 98}, 091103 (2018), 
\doi{10.1103/PhysRevD.98.091103}.

\bibitem{STAR}K. H. Ackermann et al. (STAR Collaboration), {\it STAR detector overview}, Nucl. Instrum. Methods Phys. Res. Sect. A {\bf 499}, 624 (2003), 
\doi{10.1016/S0168-9002(02)01960-5}.

\bibitem{jetrec1} B. I. Abelev et al. (STAR Collaboration), {\it Longitudinal double-spin asymmetry and cross section for inclusive neutral pion production at midrapidity in polarized proton collisions at $\sqrt{s} = 200$ GeV}, Phys. Rev. D {\bf 80}, 111108(R) (2009), 
\doi{10.1103/PhysRevD.80.111108}.

\bibitem{STAR:2019yqm}
J.~Adam \textit{et al.} (STAR Collaboration),
{\it Longitudinal double-spin asymmetry for inclusive jet and dijet production in $pp$ collisions at $\sqrt{s} = 510$  GeV},
Phys. Rev. D \textbf{100}, 052005 (2019),
\doi{10.1103/PhysRevD.100.052005}.

\bibitem{DLLMeasurement}B. I. Abelev et al. (STAR Collaboration), {\it Longitudinal spin transfer to $\Lambda$ and $\bar{\Lambda}$ hyperons in polarized proton-proton collisions at $\sqrt{s} = 200$ ${GeV}$ GeV}, Phys. Rev. D {\bf 80}, 111102(R) (2009),
 \doi{10.1103/PhysRevD.80.111102}.

\end{thebibliography}

\nolinenumbers

\end{document}